\documentclass[preprint]{rmaa}

\title{On the Size of the Non-Thermal Component in the Radio Emission from Cyg OB2 \#5}

\author{
Luis F. Rodr\'\i guez,\altaffilmark{1}
        Yolanda G\'omez,\altaffilmark{1} 
        Laurent Loinard,\altaffilmark{1}
         and Amy J. Mioduszewski\altaffilmark{2}} 
 \altaffiltext{1}{Centro de Radioastronom\'{\i}a y Astrof\'{\i}sica, Universidad
Nacional Aut\'onoma de M\'exico, Campus Morelia} 
\altaffiltext{2}{National Radio Astronomy Observatory}

\fulladdresses{
\item Yolanda G\'omez, Laurent Loinard, and Luis F. Rodr\'\i guez: Centro de Radiostronom\'{\i}a
y Astrof\'\i sica, 
UNAM, A. P. 3-72, 
(Xangari), 58089 Morelia, Michoac\'an, M\'exico
(y.gomez, l.loinard, l.rodriguez@astrosmo.unam.mx).
\item Amy J. Mioduszewski: National Radio Astronomy Observatory, 1003 Lopezville Road, Socorro, 
NM 87801, USA
(amiodusz@nrao.edu). 
}

\shortauthor{Rodr\'\i guez et al.}
\shorttitle{Size of Non-Thermal Component in Cyg OB2 \#5}

\SetVolume{50} \SetFirstPage{001} \SetYear{2010}
\ReceivedDate{\today} 
\AcceptedDate{Year Month Day} 

\resumen{
Cyg OB2 \#5 es un sistema binario de contacto con emisi\'on variable de radio.
Esta emisi\'on tiene un estado bajo de flujo dominado por el viento estelar ionizado
y un estado alto de flujo, donde una componente no-t\'ermica adicional aparece.
Ahora se sabe que las variaciones tienen un periodo
de 6.7$\pm$0.2 a\~nos. La componente no-t\'ermica ha sido atribuida a varios
agentes: una envolvente en expansi\'on eyectada periodicamente por la
binaria, emisi\'on de una regi\'on de choque de vientos, o una
estrella con emisi\'on no-t\'ermica en \'orbita exc\'entrica alrededor de
la binaria. La determinaci\'on  del tama\~no angular de la componente no-t\'ermica
es crucial para discriminar entre estas posibilidades.
Presentamos un an\'alisis de datos de archivo del VLA hechos a 
8.46 GHz en 1994 (en estado bajo)
y 1996 (en estado alto), que nos permiten sustraer el efecto de la
componente t\'ermica persistente y estimar un tama\~no angular 
$\leq 0\rlap.{''}02$
para la componente no-t\'ermica. Este tama\~no compacto favorece la explicaci\'on
en t\'erminos de una estrella con emisi\'on no-t\'ermica o
de una regi\'on de choque de vientos. 
}

\abstract{Cyg OB2 \#5 is a contact binary system with variable 
radio continuum emission. This emission has a low-flux state where
it is dominated by thermal emission from the ionized stellar
wind and a high-flux state where an additional non-thermal component appears.
The variations are now known to have 
a period of 6.7$\pm$0.2 yr. The non-thermal component has been attributed to
different agents: an expanding envelope ejected periodically from the
binary, emission from a wind-collision region,
or a star with non-thermal emission in an eccentric orbit around the
binary. The determination of the angular size of the non-thermal component
is crucial to discriminate between these alternatives. 
We present the analysis of VLA archive observations made at 8.46 GHz in 1994 (low state)
and 1996 (high state), that allow us to subtract the effect of the persistent
thermal emission and to estimate an angular size of $\leq 0\rlap.{''}02$
for the non-thermal component. This compact size favors the explanation
in terms of a star with non-thermal emission or of a
wind-collision region.
}

\keywords{STARS: INDIVIDUAL (CYG OB2 \#5) --- RADIO CONTINUUM: STARS}

\begin{document}

\maketitle

\section{Introduction}

Cyg OB2 \#5 (V729 Cyg, BD +40 4220) is an eclipsing, contact binary system consisting of
two O-type supergiants with a 6.6-day period (Hall 1974; Leung \& Schneider 1978;
Rauw et al. 1999; Linder et al. 2009). 
As several other luminous O-star systems in the Cyg OB2
association, this source was found to show evidence of 
variable radio emission (Persi et al. 1985, 1990; Bieging
et al. 1989). The radio emission appears to have
two states: a low-flux state of  $\sim$2 mJy at 4.8 GHz where the spectral index is consistent
with thermal emission from an ionized 
stellar wind, and a high-flux state of $\sim$8 mJy at 4.8 GHz
where the spectral index is flatter than in the low state. The variations were proposed to have a
7-year period (Miralles et al. 1994) and have been attributed to variable non-thermal
emission from an expanding envelope arising periodically
in the binary (Persi et al. 1990; Bieging et al.
1989; Miralles et al. 1994).

In addition, Abbott et al. (1981) and Miralles et al. (1994) reported on the existence 
of a radio ``companion'' 
0.8 arcsec to the NE of the main radio source (which is associated with the 
eclipsing binary).  Observations by Contreras et al. (1997) revealed that 
this radio source has an elongated shape and lies 
in-between the short-period binary and a third star, which was first reported by Herbig 
(1967). Contreras 
et al. (1997) suggested that the proposed NE radio companion 
actually corresponds to the wind interaction zone between 
the binary system and the tertiary component. Recently, Kennedy et al. (2010) reanalysed 
all VLA observations 
of Cyg OB2 \#5 and showed that the primary radio source, associated with the eclipsing binary,
varies with a period of  $6.7 \pm 0.2$ yr while the flux
from the secondary NE source remains constant
in time. 
These authors proposed that the variations in the main radio 
component can be represented by a simple model in which a fourth star 
(with constant non-thermal emission) moves around the eclipsing 
binary in an eccentric orbit and the varying radio emission results from 
the variable free-free opacity of the wind between the star and the observer 
(Kennedy et al. 2010). These authors estimate a major axis of 14 AU for
the star with non-thermal emission.
In this scenario, we have a quadruple star system:
the contact binary, the star with non-thermal emission
in a 6.7 yr orbit around the contact binary, and the NE component.

To advance our understanding of the nature of the
time-variable non-thermal emission, a determination
of the angular size of the source producing it is needed.  
Using VLA data taken at 8.46 GHz in 1991 October 3 (when Cyg OB2 \#5 was in the high radio
state), Miralles et al. (1994) estimated an angular size of 
$0\rlap.{''}046 \pm 0\rlap.{''}006$ for the whole (non-thermal plus thermal components)
emission. Using MERLIN data taken at
4.8 GHz in 1996 November 14 (when
Cyg OB2 \#5 was again in the high radio
state), Kennedy et al. (2010) estimated an
angular size of $\sim0\rlap.{''}077$
for the whole (non-thermal plus thermal components)
emission associated with the primary radio component. 
Since the thermal contribution comes from the ionized
wind that is known to be extended, these size estimates can be considered as upper limits
to the size of the non-thermal emission. In this paper
we present the analysis of archive 8.46 GHz 
continuum observations made with the Very
Large Array (VLA)
of the NRAO\footnote{The National Radio 
Astronomy Observatory is operated by Associated Universities 
Inc. under cooperative agreement with the National Science Foundation.} 
in the A configuration toward Cyg OB2 \#5 in two
epochs. During 1994 April 9 (1994.27), the source was in the low radio state
and only the thermal emission from the ionized stellar wind was present.
These observations were used to estimate the characteristics 
of the persistent thermal component. During 1996 December 28 (1996.99), the source was in the 
high radio state, with both the thermal and non-thermal components
present. Using the information obtained from the 1994 data, it is 
possible to subtract in the \sl(u,v) \rm plane the contribution from the thermal component 
and obtain a more stringent estimate of the angular size of 
the non-thermal emission. 

\section{Data Reduction}

The archive data
were edited and calibrated using the software package Astronomical Image
Processing System (AIPS) of NRAO. The parameters of
the observations are given in Table 1. The absolute amplitude
calbrator for both epochs was 1331+305, with an
adopted flux density of 5.21 Jy. The \sl (u,v) \rm data were self-calibrated in amplitude
and phase. An image was made with the 1994.27 \sl (u,v) \rm data and the clean components 
of this image were
subtracted from the 1996.99 \sl (u,v) \rm data using the AIPS task UVSUB.
The resulting \sl (u,v) \rm data, that we refer to as 1996.99-1994.27,
was analyzed both in the image and in the \sl (u,v) \rm planes. In the image
plane a compact, basically unresolved source is observed. A Gaussian
ellipsoid fit to this
source in the image plane using the AIPS task JMFIT indicates maximum possible
deconvolved
angular dimensions of $0\rlap.{''}035 \times 0\rlap.{''}029$ 
(major axis $\times$ minor axis) for the source. 
However, White \& Becker (1982) have shown that it is better to study the
angular dimensions of marginally
resolved sources directly in the \sl (u,v) \rm plane.
We then fitted directly the \sl (u,v) \rm data in two dimensions
using the AIPS task UVFIT
and a Gaussian ellipsoid model. The fit gave $0\rlap.{''}023\pm0\rlap.{''}013 
\times 0\rlap.{''}023\pm0\rlap.{''}019$
(major axis $\times$ minor axis) for the source.
This result indicates that the source is very compact and
that, within the error, it
shows no strong departures from a circular morphology. 
We then adopt as an estimate of the Gaussian angular size (full width
at half maximum) of the
source the geometric mean of the major and minor axes, that
gives $\theta_G = 0\rlap.{''}023\pm0\rlap.{''}012$.

\begin{table*}[htbp]
\small
 \setlength{\tabnotewidth}{2.0\columnwidth} 
 \tablecols{7} 
 \caption{Archive Data Used}
 \begin{center}
   \begin{tabular}{lcccccc}\hline\hline
                     &         & Frequency & On-source & Phase & Bootstrapped & Beam  \\
Epoch                 & Project & (GHz)& Time (min) & 
Calibrator & Flux Density (Jy) & Angular Size $^a$ \\ 
\hline
1994 Apr 09 (1994.27) & AR277   & 8.46 & 37 & 2007+404 & 2.84$\pm$0.01 &
$0\rlap.{''}18 \times 0\rlap.{''}16;~\ $-$43^\circ$  \\
1996 Dec 28 (1996.99) & AR277   & 8.46 & 45 & 2007+404 & 2.89$\pm$0.01 &
$0\rlap.{''}30 \times 0\rlap.{''}21;~\ $-$85^\circ$  \\
\hline\hline
\tabnotetext{a}{Major axis $\times$ minor axis; position angle, for a 
\sl (u,v) \rm weighting of ROBUST = 0 (Briggs 1995).}
   \label{tab:1}
   \end{tabular}
 \end{center}
\end{table*}

The angular size determined for the non-thermal emission depends on the
model assumed to fit the \sl (u,v) \rm data. 
A model of a circular 
disk with constant brightness
gives an angular diameter of $\theta_D = 0\rlap.{''}019\pm0\rlap.{''}010$. 
Given that the errors are comparable to the measurements, we will 
take this determination as an upper limit, $\theta('') \leq 0\rlap.{''}02$,
to the angular size of the non-thermal emission.

\section{Interpretation}

The small angular size found for the non-thermal component,
$\theta('') \leq 0\rlap.{''}02$,
can be used to discuss the nature of the
non-thermal component. An expanding envelope ejected periodically
from the binary (Persi et al. 1990; Bieging et al.
1989; Miralles et al. 1994) would have to be bigger than the
radio photosphere to be outside the optically thick region
around the contact binary system. We can use the results
for the 1994 data discussed in the Appendix
to estimate that the radio photosphere
at 8.46 GHz has an angular size of $\theta('') \simeq 72$ milli-arcseconds.
We then conclude that an envelope cannot explain the observations since
the non-thermal emission 
would be significantly larger than observed.
Recently, Pittard (2010) has proposed that the time variable emission
observed in O+O binaries can come from the thermal emission from the
wind-collision region between the stars. In the case of Cyg OB2 \#5
the wind-collision region between the members of the contact binary
is probably a strong emitter, but lies deep inside the radio photosphere
of the system, although it is known that
the radio photospheres can be clumpy or asymmetric,
allowing the detection of sources embedded
in it (e.g. WR140; White \& Becker 1991). The wind-collision 
region could also be located between
the contact binary and a third star. Kennedy et al. (2010) discussed
the variations arising from non-thermal
emission in a wind-collision region between the contact binary and a third star. They
demonstrate that thermal emission variations that may arise in such a wind-collision region 
will not have sufficient amplitude to account for the variations,
and conclude that the thermal-emission Pittard models
would predict smaller flux changes than observed.
The Kennedy et al. (2010) models also account well for the variable radio emission.
We then favor the possibility of a third star with non-thermal emission
in orbit around the contact binary, as proposed by Kennedy et al. (2010),
or a wind-collision region.

Even when the non-thermal emission from young,
low-mass stars is typically associated with
gyrosynchrotron emission from Lorentz $\gamma$ factors of a few
(e.g. Ray et al. 1997), while the emission from massive stars is
believed to be synchrotron from relativistic electrons with
very large $\gamma$ factors (e.g. Pittard \& Dougherty, 2006), we
cannot rule out the possibility that the non-thermal emission from
Cyg OB2 \#5 is being produced by a young,
low-mass star (i.e. a  Tauri star) that formed coevally with the rest of the system.
The non-thermal emission from young,
low-mass stars is usually very compact.
For the T Tau Sb star, with a mass of 0.6 $M_\odot$, Loinard et al. (2005; 2007) estimate an
angular size of 0.5 mas, that at a distance of 147.6 pc
corresponds to a radius of $\sim$8 $R_\odot$. 
For the more massive S1 star (6 $M_\odot$) in Ophiuchus, Loinard et al. (2008)
estimate an angular size of 0.95 mas, that at a distance of 116.9 pc
corresponds to a radius of $\sim$12 $R_\odot$.
At the larger distance of 925 pc (Linder et al. 2009), we expect
a non-thermal star associated with Cyg OB2 \#5 to have an angular size
well below the milli-arcsecond.

Cyg OB2 \#5 will be again in high state during 2010 and 2011
and a very long baseline interferometry study is needed to determine accurately
the angular size and morphology of the compact non-thermal emission. 

\section{Conclusions}

We present the analysis of VLA archive data of the Cyg OB2 \#5 system
taken during the high (1996) and low (1994) radio states. This analysis
allows the subtraction, in the \sl (u,v) \rm plane of the persistent
thermal component and a better estimate of the angular dimensions of the
non-thermal component. We obtain
an upper limit, $\theta('') \leq 0\rlap.{''}02$
for the angular size of the non-thermal emission.
Cyg OB2 \#5 will be again in high state during 2010 and 2011
and a very long baseline interferometry study is needed to determine
the angular size and morphology of the compact non-thermal emission.

\acknowledgments
We are thankful for the support
of DGAPA, UNAM, and of CONACyT (M\'exico).
This research has made use of the SIMBAD database, 
operated at CDS, Strasbourg, France.

\vskip0.5cm

\centerline{APPENDIX}

\begin{appendix}

\section{One-dimensional Analysis of the Visibility Function}

\begin{figure}
\centering
\includegraphics[scale=0.44, angle=0]{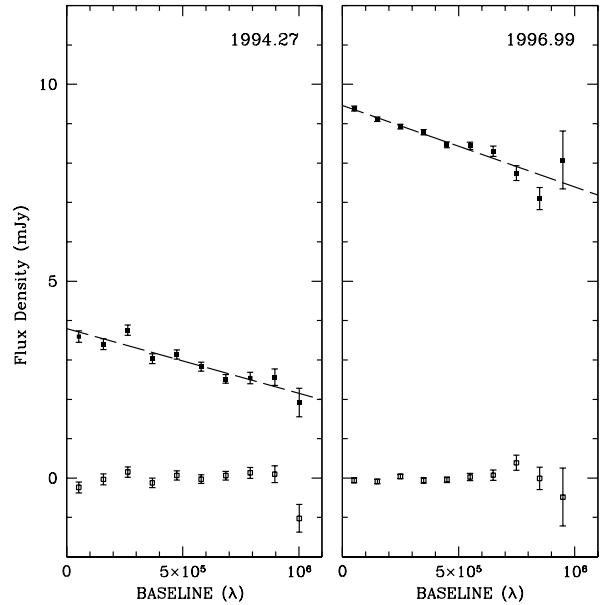}
\caption{(Left) Real (filled squares) and imaginary (empty squares) components of the
emission at 8.46 GHz for 1994 April 9. The dashed line is the least squares fit
to the real component. (Right) Same as in the left panel for
1996 December 28. The imaginary component is consistent with zero, indicating that
the source is symmetric about the phase center (the
origin of the visibility plane) and has no significant structure on these spatial
scales.
}
 \label{fig1}
\end{figure}

If the sources have approximate circular symmetry, the analysis of the
\sl (u,v) \rm data can be made in a one-dimension approximation.
To test if this was possible with the data discussed here,
the NE component was subtracted in
two dimensions in the \sl (u,v) \rm plane of both data sets 
using Gaussian ellipsoid models to leave
only the emission from the main component, the one directly associated
with the contact binary. This subtraction was performed using the AIPS
task UVMOD. We then fitted the \sl (u,v) \rm data in
two dimensions using the AIPS task UVFIT
and a Gaussian ellipsoid model. The fits gave $0\rlap.{''}079\pm0\rlap.{''}007 
\times 0\rlap.{''}066\pm0\rlap.{''}008$
(major axis $\times$ minor axis) for the 1994.27 data
and $0\rlap.{''}045\pm0\rlap.{''}006 
\times 0\rlap.{''}036\pm0\rlap.{''}005$
(major axis $\times$ minor axis) for the 1996.99 data.
These results indicate that,
within error, the emission at both epochs does not 
show strong departures from a circular assumption. 
We also note that at a distance of 925 pc (Linder et al. 2009),
the semimajor axis of the non-thermal star (7 AU) will subtend an
angle of only 0.008", below the upper limit of 0.020" found for the
angular size of the non-thermal component. It is quite possible that this 
non-thermal star, if present, will
introduce asymmetries but future VLBI observations will be needed to
test this.

In Figure 1
we present the one-dimensional real and imaginary components of the amplitude as a
function of baseline. These values were obtained using the AIPS
task UVPLT that averages in circular rings around the origin
of the \sl (u,v) \rm plane. Before applying this task, the source emission
was centered in this origin.

For a marginally resolved wind source,
the real part of the visibility, $V(b)$, can be fitted in
one-dimension with
a linear function (Escalante et al. 1989):

$$V(b) = S_0  - A b, \eqno(1)$$

\noindent where $S_0$ is the flux density at zero spacing (the total flux density),
given in mJy, $A$ is the slope of the linear fit, and $b$ is the baseline separation
in wavelengths. For 1994 only the thermal (T) emission was present
and we obtain $S_0(T) = 3.79 \pm 0.09$ mJy and
$A(T) = (1.64 \pm 0.18) \times 10^{-6}$ mJy wavelengths$^{-1}$.
For 1996 both the thermal and non-thermal (T+N) emissions
were present and we obtain $S_0(T+N) = 9.46 \pm 0.05$ mJy and
$A(T+N) = (2.07 \pm 0.14) \times 10^{-6}$ mJy wavelengths$^{-1}$.
The fact that both epochs give similar slopes suggests that,
as expected, most of the extended emission comes from the thermal
component.
Subtracting the 1994 fit to the 1996 fit we obtain the
non-thermal (N) contribution alone, that is given
by $S_0(N) = 5.67 \pm 0.10$ mJy and
$A(N) = (0.43 \pm 0.23) \times 10^{-6}$ mJy wavelengths$^{-1}$.
The characteristic angular size of the emitting region is given by
(Escalante et al. 1989; Miralles et al. 1994):

$$\theta('') \simeq 1.44 \times 10^5~{{A} \over {S_0}}, \eqno(2)$$

\noindent which gives $\theta('') \simeq 0\rlap.{''}011 \pm 0\rlap.{''}006$ for the non-thermal
component. 
It should be noted that our fitting of the non-thermal
component to a thermal wind model may not be
consistent since the structure of the visibilities for an optically
thin non-thermal source and a partially optically thick stellar wind are
not necessarily the same.
In Table 2 we summarize the parameters obtained for the thermal and
thermal plus non-thermal states as well as the subtraction, that
traces the non-thermal emission alone.

\begin{table}[htbp]
\small
 \setlength{\tabnotewidth}{0.8\columnwidth} 
 \tablecols{4} 
 \caption{Parameters from fits}
 \begin{center}
   \begin{tabular}{lccc}\hline\hline
  &  $S_0$   & $A$ & $\theta$ \\
Emission$^a$  & (mJy) & (10$^{-6}$ mJy $\lambda^{-1}$)& (mas$^b$) \\
\hline
T & 3.79$\pm$0.09 & 1.64$\pm$0.18 & 62$\pm$7 \\
N+T & 9.46$\pm$0.05 & 2.07$\pm$0.14 & 32$\pm$2 \\
N & 5.67$\pm$0.10 & 0.43$\pm$0.23 & 11$\pm$6 \\
\hline\hline
\tabnotetext{a}{T = Thermal, N = Non-thermal.}
\tabnotetext{b}{mas = milli-arcseconds.}
   \label{tab:2}
   \end{tabular}
 \end{center}
\end{table}

\end{appendix}


\end{document}